\documentclass[aps,prd,showpacs,twocolumn]{revtex4}

\usepackage{epsfig}
\usepackage{longtable}
\usepackage{morefloats}

\begin{document}

\title{Properties of Excited Charm and Charm-Strange Mesons} 
\author{ Stephen Godfrey\footnote{Email: godfrey@physics.carleton.ca} and Kenneth Moats}
\affiliation{
Ottawa-Carleton Institute for Physics, 
Department of Physics, Carleton University, Ottawa, Canada K1S 5B6 
}

\date{January 15, 2016}

\begin{abstract}
We calculate the properties of excited charm and charm-strange mesons.  We use the
relativized quark model to calculate their masses and wavefunctions that are used
to calculate radiative transition partial widths and the $^3P_0$ quark-pair-creation model
to calculate their strong decay widths.  We use these results to make quark model
spectroscopic assignments for recently observed charm and charm-strange mesons.  In particular
we find that the properties of the $D_J(2550)^0$ and $D_J^*(2600)^0$ are consistent 
with those of the $2^1S_0(c\bar{u})$ and the  $2^3S_1(c\bar{u})$ states respectively, 
the $D_1^*(2760)^0$, $D_3^*(2760)^-$, and $D_J(2750)^0$ with those of the 
$1^3D_1(c\bar{u})$, $1^3D_3(d\bar{c})$, and $1D_2(c\bar{u})$ states respectively.
We tentatively identify the $D^*_J(3000)^0$ as the $1^3F_4(c\bar{u})$ and favour
the  $D_J(3000)^0$ to be the $3^1S_0(c\bar{u})$ although we do not rule out the $1F_3$
and $1F_3'$ assignment.  For the recently observed charm-strange mesons we identify the
$D_{s1}^*(2709)^\pm$, $D_{s1}^*(2860)^-$, and $D_{s3}^*(2860)^-$ as
the $2^3S_1(c\bar{s})$, $1^3D_1(s\bar{c})$, and $1^3D_3(s\bar{c})$ states respectively and suggest
that the $D_{sJ}(3044)^\pm$ is most likely the $D_{s1}(2P_1')$ or  $D_{s1}(2P_1)$ states
although it might be the $D_{s2}^*(2^3P_2)$ with the $DK$ final state  too small
to be observed with current statistics.  Based on the predicted properties of excited
states, that they not have too large a total width and they have a reasonable branching 
ratio to simple final states, we suggest states that should be able to be found in the 
near future. We expect that the tables of properties summarizing our results will be useful for interpreting 
future observations of charm and charm-strange mesons.

\end{abstract}
\pacs{12.39.Pn, 13.25.-k, 13.25.Ft, 14.40.Lb}

\maketitle

\section{Introduction}

Over the last decade, charm meson spectroscopy has undergone a resurgence due to the discovery
of numerous excited charm and charm-strange states by the $B$-Factory experiments 
BaBar and Belle 
\cite{Bevan:2014iga,delAmoSanchez:2010vq,Aubert:2006mh,Brodzicka:2007aa,Aubert:2009ah,Lees:2014abp,Olive:2014kda}
and by the CLEO experiment \cite{Besson:2003cp}.  
More recently the LHCb experiment has 
demonstrated the capability of both observing these states and determining their properties
\cite{Aaij:2012pc,Aaij:2013sza,Aaij:2014baa,Aaij:2015vea,Aaij:2014xza,Aaij:2015sqa}.
This has led to considerable theoretical interest in attempting to make quark model spectroscopic
assignments for these new states
 by comparing theoretical predictions to experimental measurements
\cite{Zhang:2006yj,Close:2006gr,Wang:2010ydc,Xiao:2014ura,Chen:2011rr,Badalian:2011tb,Wang:2013tka,Godfrey:2013aaa,Sun:2013qca,Godfrey:2014fga,Song:2014mha,Ke:2014ega,Wang:2014jua,Yu:2014dda,Li:2010vx,Colangelo:2006rq,vanBeveren:2009jq,Song:2015fha}.
At the same time, steady progress is being made in lattice QCD \cite{Moir:2013ub,Lewis:2000sv,Cichy:2015tma}
for which these experimental results
and spectroscopic classifications are an important benchmark. 
With the start of higher energy and higher luminosity beams at the LHC and higher luminosity 
at the SuperKEKB $e^+e^-$ collider we expect that more new states will be observed.  To 
identify newly discovered states, a theoretical roadmap is needed.  The quark model
has been successful in taking on this role and we turn to it to calculate the properties 
of excited charm and charm-strange mesons. 

An important property of heavy-light mesons is that in the limit that the heavy quark
mass becomes infinite the properties of the meson are determined by those of the light
quark \cite{DeRujula:1976kk,Rosner:1985dx,isgur91}.  The light quarks are characterized
by their total angular momentum $j_q$ such that 
$\vec{j}_q=\vec{s}_q+\vec{L}$ where ${s}_q$ 
is the light quark spin and ${L}$ is
its orbital angular momentum.  $j_q$ is combined with $S_Q$, the spin of the heavy quark, to give
the total angular momentum of the meson.  The quantum numbers $S_Q$ and $j_q$ are 
separately conserved.  Thus, for a given $L$, the states will be grouped into doublets
characterized by the angular momentum of the light quark.  
For example, the four $L=1$ $P$-wave mesons can be grouped into two doublets
characterized by the angular momentum of the light quark $j_q=3/2$ with $J^P=1^+, \; 2^+$
and $j_q=1/2$ with $J^P=0^+, \; 1^+$ where $J$ and $P$ are the total angular momentum and parity 
of the excited meson.  In the heavy quark limit (HQL) the members of the doublets will be degenerate
in mass, and this degeneracy is broken by $1/m_Q$ corrections \cite{isgur91,eichten93}.  For the $L=1$ 
multiplet, heavy quark
symmetry and conservation of parity and $j_q$ also predict that the strong decays 
$D_{(s)J}^{(*)} (j_q =3/2) \to D^{(*)}\pi (K) $ will only proceed through a $D$-wave while 
the decays $D_{(s)J}^{(*)} (j_q =1/2) \to D^{(*)}\pi (K) $ will only proceed via an $S$-wave
\cite{Godfrey:1986wj,Lu:1991px}. The states decaying to a $D$-wave are expected to be
narrow due to the angular momentum barrier while those decaying to an $S$-wave are expected 
to be broad.  Similar patterns are predicted for higher  $L$ multiplets so that measuring 
the properties of excited charm mesons can be used to both help identify them and to 
see how well excited states are described by the properties expected in the heavy quark limit.
However, for higher mass states more phase space is available, leading to more possible decay  
channels, resulting in more complicated decay patterns so that the predictions of the HQL are
less apparent.

Our goals for this paper are twofold.  First  we want to provide 
a roadmap of charm and charm-strange meson properties
to identify which states are the most promising 
candidates to be observed and the final states they are most likely to be observed in. 
Hereafter, for conciseness, we will generally refer to both 
charm and charm-strange mesons as charm mesons.
Second, when new states are observed we can use our roadmap 
to make quark model spectroscopic assignments for these newly found states.

In the first part of this paper we calculate the masses and wavefunctions
of excited charm and charm-strange mesons using the 
relativized quark model \cite{Godfrey:1985xj} which we describe in the next section.
 Radiative
transitions are described in Section~\ref{sec:emtransitions}  
and strong decay widths are calculated using the $^3P_0$ quark-pair 
creation model \cite{Micu:1968mk,Le Yaouanc:1972ae} which is
described in Section~\ref{sec:strongdecays}.
These models have been described 
extensively in the literature so rather than repeating detailed descriptions of these models
we will give brief summaries and refer the interested reader to the references  
for further details.  The outcome of this part of the paper is a comprehensive summary of
excited charm meson properties.

In the second part, in Sections~\ref{sec:charm_classification} and \ref{sec:charm_strange_classification}, 
we  use these results to examine the numerous newly observed charm and charm-strange mesons and 
attempt to make quark model spectroscopic assignments. 
This approach has been used in numerous papers although in some cases different 
calculations come to different conclusions.  Thus, another goal of this paper is to 
suggest further diagnostic measurements that can resolve these differences to give 
an unambiguous spectroscopic assignment.  
In Section~\ref{sec:finding_the_missing} we will
use the quark model roadmap we produced in the first part of this paper 
 to suggest which missing states, because of their properties,
are most
likely to be observed in the near future and suggest the most promising final states
to study. 
We summarize our conclusions in the final section.

\section{Spectroscopy}
\label{sec:spectroscopy}

We use the relativized quark model \cite{Godfrey:1985xj} 
(see also Ref.~\cite{Godfrey:1985by,godfrey85b,Godfrey:2004ya,Godfrey:2005ww})
to calculate meson masses and their wavefunctions which we use to calculate 
decay properties.  
The model is described in detail in Ref.~\cite{Godfrey:1985xj} to which we direct the interested
reader.  The general characteristics of this model are that it assumes 
a relativistic kinetic energy term and the potential
incorporates a Lorentz vector one-gluon-exchange interaction with a
QCD motivated running coupling constant, $\alpha_{s}(r)$, and a  Lorentz scalar linear confining 
interaction.
This is typical of most such models which 
are based on some variant of the Coulomb plus 
linear potential expected from QCD and that often include some relativistic effects
\cite{Gupta:1994mw,Ebert:1997nk,DiPierro:2001uu,Close:2005se,Ebert:2009ua,Radford:2009bs,Shah:2014caa,Song:2015fha}.  
The relativized quark 
model has been reasonably successful in describing most known mesons and has proven to be
a useful guide to understanding  newly found states
\cite{Barnes:2005pb,Godfrey:2014fga,Barnes:2003vb,Godfrey:2003kg,Blundell:1995ev,Godfrey:1986wj}.
However in recent years, starting with the discovery of the 
$D_{sJ}(2317)$ \cite{Aubert:2003fg,Besson:2003cp,Krokovny:2003zq} and $X(3872)$ states \cite{Choi:2003ue},
an increasing number of states have been observed that do not fit into this picture 
\cite{Godfrey:2008nc,Godfrey:2009qe,Braaten:2013oba,DeFazio:2012sg}  
pointing to the need to include physics which has hitherto been neglected such as 
coupled channel effects \cite{Eichten:2004uh} which appears to be most important for 
states lying near kinematic thresholds.  As a consequence of neglecting coupled channel effects
and the crudeness of the relativization procedure we do not 
expect the mass predictions to be accurate to better than $\sim 10-20$~MeV.

For the case of a quark and antiquark of unequal mass, charge conjugation
parity is no longer a good quantum number so that states with different 
total spins but with the same total angular momentum, such as
the $^3P_1 -^1P_1$ and $^3D_2 -^1D_2$ pairs, can mix via
the spin orbit interaction or some other mechanism.
Consequently, the physical $J=1$ $P$-wave states are linear
combinations of $^3P_1$ and $^1P_1$ which we describe by:
\begin{eqnarray}
\label{eqn:mixing}
P  & = &  {^1}P_1 \cos\theta_{nP} + {^3}P_1 \sin\theta_{nP} \nonumber \\   
P' & = & -^1P_1 \sin\theta_{nP} + {^3}P_1 \cos \theta_{nP} 
\end{eqnarray}
where 
$P\equiv L=1$ designates the relative angular momentum of the $c\bar{q}$ 
pair and the subscript $J=1$ is the total angular momentum of the $c\bar{q}$ 
pair which is equal to $L$, and $q$ can represent either a $u$, $d$ or $s$ quark.  
There are analogous expressions for higher $L$ states where  $L=D$, $F$, etc.
Our notation implicitly implies $L-S$ 
coupling between the quark spins and the relative orbital angular momentum.  
In the heavy quark limit in which 
the heavy quark mass $m_Q\to \infty$, 
the states can be described by the total angular momentum of the
light quark, $j_q$, which couples to the spin of the heavy quark and
corresponds to $j-j$ coupling.  In this limit the mixed states 
are given by \cite{Cahn:2003cw}
\begin{eqnarray}
\label{eqn:hqlmixing}
|J= L, j_q=L + \frac{1}{2} \rangle & = & \sqrt{{J+1}\over{2J+1}} | J=L, S=0 \rangle \nonumber \\
			& &  + \sqrt{\frac{J}{2J+1}} | J=L, S=1 \rangle \nonumber \\
|J= L, j_q=L - \frac{1}{2} \rangle & = & - \sqrt{{J}\over{2J+1}} | J=L, S=0 \rangle \nonumber \\
			& &  + \sqrt{{J+1}\over{2J+1}} | J=L, S=1 \rangle
\end{eqnarray}
The $j_q=L-\frac{1}{2}$ state that is mainly spin triplet corresponds to the primed state in 
eqn.~\ref{eqn:mixing} and the $j_q=L+\frac{1}{2}$ that is mainly spin singlet corresponds
to the unprimed state.
For $L=1$ the HQL gives rise to two doublets, one with $j_q=1/2$ and the 
other with  $j_q=3/2$ and with the conventions of eqns.~\ref{eqn:mixing} and \ref{eqn:hqlmixing}
corresponds to $\theta_P=\tan^{-1}(-1/\sqrt{2})\simeq -35.3^\circ$.
For $L=2$ the HQL gives  two doublets with $j_q=3/2 $ and $5/2$ 
with mixing angle $\theta_D=-\tan^{-1}(\sqrt{2/3})=-39.2^\circ$. The minus signs arise from
our $c\bar{q}$ convention.  
Some 
authors prefer to use the $j-j$ basis \cite{eichten94}
but since we solve our 
Hamiltonian equations assuming $L-S$ eigenstates and then include the 
$LS$ mixing we use the notation of eqn. \ref{eqn:mixing}.  
Radiative transitions are  sensitive to the 
$^3L_L-^1L_L$ mixing angle.
We note that the 
definition of the mixing angles are fraught with ambiguities.  For 
example, charge conjugating $c\bar{q}$ into $q\bar{c}$ flips the 
sign of the angle and the phase convention depends on the
order of coupling $\vec{L}$, $\vec{S}_q$ and $\vec{S}_{\bar{q}}$
\cite{barnes}.

\begin{figure*}[th]
\begin{center}
\includegraphics[width=6.0in]{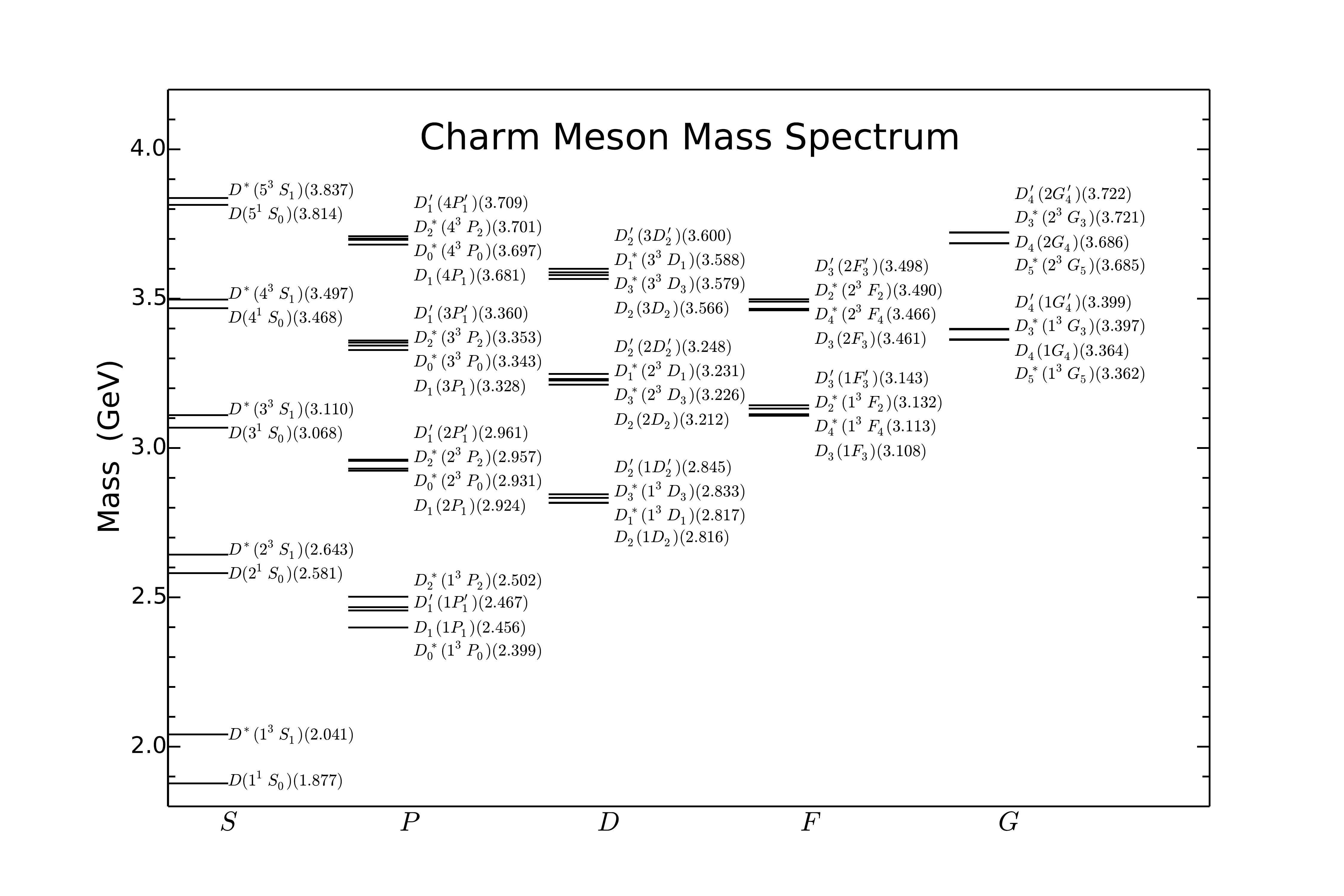}
\end{center}
\caption{The charm meson mass spectrum as predicted by the relativized quark model \cite{Godfrey:1985xj}.
The $^3L_L-^1L_L$ mixing angles are given in Tables \ref{tab:SPmasses} and \ref{tab:DFGmasses}.
}
\label{Fig1}
\end{figure*}

\begin{figure*}[th]
\begin{center}
\includegraphics[width=6.0in]{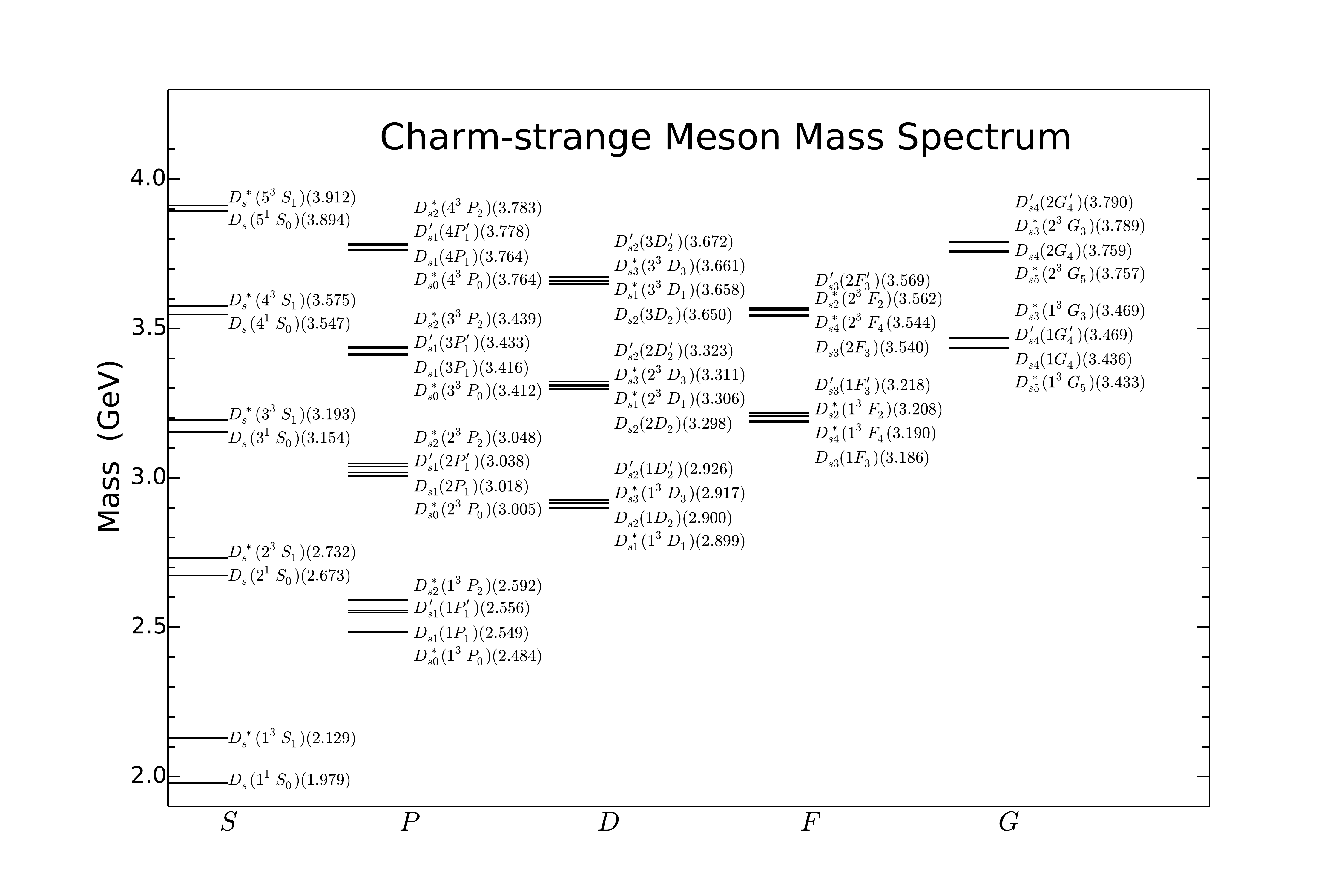}
\end{center}
\caption{The charm-strange meson mass spectrum as predicted by the relativized quark model \cite{Godfrey:1985xj}.
The $^3L_L-^1L_L$ mixing angles are given in Tables \ref{tab:SPmasses} and \ref{tab:DFGmasses}.
}
\label{Fig2}
\end{figure*}

To solve the Hamiltonian to obtain masses and wavefunctions we used the following parameters: the slope of the
linear confining potential is 0.18 GeV$^2$, 
$m_q=0.22$ GeV, $m_s=0.419$~GeV and $m_c=1.628$ GeV.  The predictions of our model for 
the charm mesons are given in Fig.~\ref{Fig1} and for the charm-strange mesons in Fig~\ref{Fig2}
and the predicted masses and $^3L_L - ^1L_L$ mixing angles are given in
Tables~\ref{tab:SPmasses} and \ref{tab:DFGmasses}.

\begin{table}[th]
\caption{Predicted charm and charm-strange $S$ and $P$-wave meson masses, spin-orbit mixing angles
and $\beta_{eff}$'s.  
The $P_1 -P_1'$ states and mixing angles 
are defined using the convention of eqn. \ref{eqn:mixing}. 
Where two values of $\beta_{eff}$ are listed, the first  value is for the singlet state
and the second value is for the triplet state.
\label{tab:SPmasses}}
\begin{tabular}{lllll} \hline \hline
State 
			& \multicolumn{2}{c}{$c\bar{q}$}  & \multicolumn{2}{c}{$c\bar{s}$}  \\ \hline
			& Mass & $\beta_{eff}$ 	& Mass & $\beta_{eff}$ \\ \hline
$1^3S_1 $ 	& 2041 & 0.516 		& 2129 & 0.562 \\
$1^1S_0 $ 	& 1877 & 0.601			& 1979 & 0.651 \\ \hline
$2^3S_1 $ 	& 2643 & 0.434 		& 2732 & 0.458 \\
$2^1S_0 $ 	& 2581 & 0.450 		& 2673 & 0.475 \\ \hline
$3^3S_1 $ 	& 3110 & 0.399 		& 3193 & 0.415 \\
$3^1S_0 $ 	& 3068 & 0.407 		& 3154 & 0.424 \\ \hline
$4^3S_1 $ 	& 3497 & 0.382 		& 3575 & 0.393 \\
$4^1S_0 $ 	& 3468 & 0.387 		& 3547 & 0.400 \\ \hline
$5^3S_1 $ 	& 3837 & 0.371 		& 3912 & 0.383 \\
$5^1S_0 $ 	& 3814 & 0.376 		& 3894 & 0.393 \\ \hline
$1^3P_2 $ 	& 2502 & 0.437 		& 2592 & 0.464 \\
$1 P_1 $ 		& 2456 & 0.475, 0.482 	& 2549 & 0.498, 0.505 \\
$1 P_1' $  	& 2467 & 0.475, 0.482 	& 2556 & 0.498, 0.505 \\
$1^3P_0 $ 	& 2399 & 0.516 		& 2484 & 0.542 \\
$\theta_{1P}$ 	& -25.68$^\circ$ & 		& -37.48$^\circ$ & \\ \hline
$2^3P_2 $ 	& 2957 & 0.402 		& 3048 & 0.420\\
$2 P_1 $ 		& 2924 & 0.417, 0.419 	& 3018 & 0.433, 0.434 \\
$2 P_1' $  	& 2961 & 0.417, 0.419 	& 3038 & 0.433, 0.434 \\
$2^3P_0 $ 	& 2931 & 0.431 		& 3005 & 0.444\\ 
$\theta_{2P}$ 	& -29.39$^\circ$ & 		& -30.40$^\circ$ & \\ \hline
$3^3P_2 $ 	& 3353 & 0.383			& 3439 & 0.396 \\
$3 P_1 $ 		& 3328 & 0.392, 0.392 	& 3416 & 0.404, 0.404 \\
$3 P_1' $  	& 3360 & 0.392, 0.392  	& 3433 & 0.404, 0.404\\
$3^3P_0 $ 	& 3343 & 0.398 		& 3412 & 0.409 \\
$\theta_{3P}$ 	& -28.10$^\circ$ & 		& -27.72$^\circ$ & \\ \hline
$4^3P_2 $ 	& 3701 & 0.371 		& 3783 & 0.382 \\
$4 P_1 $ 		& 3681 & 0.378, 0.377 	& 3764 & 0.387, 0.387 \\
$4 P_1' $  	& 3709 & 0.378, 0.377 	& 3778 & 0.387, 0.387 \\
$4^3P_0 $ 	& 3697 & 0.381 		& 3764 & 0.390 \\
$\theta_{4P}$ 	& -26.91$^\circ$ & 		& -25.43$^\circ$ & \\
\hline
\hline
\end{tabular}
\end{table}

\begin{table}[th]
\caption{Predicted charm and charm-strange $D$, $F$, and $G$-wave meson masses, spin-orbit mixing angles
and $\beta_{eff}$'s.  
The  $D_2 -D_2'$,  $F_3 -F_3'$ and $G_4 -G_4'$ states and mixing angles 
are defined using the convention of eqn. \ref{eqn:mixing}. 
Where two values of $\beta_{eff}$ are listed, the first  value is for the singlet state
and the second value is for the triplet state.
\label{tab:DFGmasses}}
\begin{tabular}{lllll} \hline \hline
State 
			& \multicolumn{2}{c}{$c\bar{q}$}  & \multicolumn{2}{c}{$c\bar{s}$}  \\ \hline
			& Mass & $\beta_{eff}$ 	& Mass & $\beta_{eff}$ \\ \hline
$1^3D_3 $ 	& 2833 & 0.407			& 2917 & 0.426 \\
$1 D_2 $ 		& 2816 & 0.428, 0.433	& 2900 & 0.444, 0.448 \\
$1 D_2' $  	& 2845 & 0.428, 0.433	& 2926 & 0.444, 0.448 \\
$1^3D_1 $ 	& 2817 & 0.456			& 2899 & 0.469  \\
$\theta_{1D}$ 	& -38.17$^\circ$ &  		& -38.47$^\circ$ & \\    \hline
$2^3D_3 $ 	& 3226 & 0.385	 	& 3311 & 0.400 \\
$2 D_2 $ 		& 3212 & 	0.396, 0.399	& 3298 & 0.408, 0.410 \\
$2 D_2' $  	& 3248 & 	0.396, 0.399	& 3323 & 0.408, 0.410\\
$2^3D_1 $ 	& 3231 & 	0.410		& 3306 & 0.419 \\
$\theta_{2D}$ 	& -37.44$^\circ$ &  		& -37.71$^\circ$ & \\    \hline
$3^3D_3 $ 	& 3579 & 0.372			& 3661 & 0.383 \\
$3 D_2 $ 		& 3566 & 0.379, 0.381	& 3650 & 0.389, 0.390\\
$3 D_2' $  	& 3600 & 0.379, 0.381	& 3672 & 0.389, 0.390\\
$3^3D_1 $ 	& 3588 & 0.387			& 3658 &  0.395\\
$\theta_{3D}$ 	& -36.90$^\circ$ &  		& -37.15$^\circ$ & \\    \hline
$1^3F_4 $ 	& 3113 &  	0.390		& 3190 & 0.405 \\
$1 F_3 $ 		& 3108 &  	0.404, 0.407	& 3186 & 0.417, 0.419 \\
$1 F_3' $  		& 3143 &  	0.404, 0.407	& 3218 & 0.417, 0.419 \\
$1^3F_2 $ 	& 3132 &  	0.423		& 3208 & 0.432 \\
$\theta_{1F}$ 	& -39.52$^\circ$ & 		& -39.30$^\circ$ & \\ \hline
$2^3F_4 $ 	& 3466 &  	0.374		& 3544 & 0.386 \\
$2 F_3 $ 		& 3461 &  	0.383, 0.385	& 3540 & 0.393, 0.394 \\
$2 F_3' $  		& 3498 &  	0.383, 0.385	& 3569 & 0.393, 0.394 \\
$2^3F_2 $ 	& 3490 &  	0.394		& 3562 & 0.401 \\
$\theta_{1F}$ 	& -39.38$^\circ$ & 		& -39.12$^\circ$ & \\ \hline
$1^3G_5 $ 	& 3362 &  0.379		& 3433 & 0.391  \\
$1 G_4 $ 		& 3364 &  	0.389, 0.391	& 3436 & 0.399, 0.401  \\
$1 G_4' $  	& 3399 &  	0.389, 0.391	& 3469 & 0.399, 0.401  \\
$1^3G_3 $ 	& 3397 &  	0.402		& 3469 & 0.410 \\
$\theta_{1G}$ 	& -40.18$^\circ$ & 		& -39.95$^\circ$ & \\ \hline
$2^3G_5 $ 	& 3685 &  	0.367		& 3757 & 0.377 \\
$2 G_4 $ 		& 3686 &  	0.374, 0.375	& 3759 & 0.382, 0.383 \\
$2 G_4' $  	& 3722 &  	0.374, 0.375	& 3790 & 0.382, 0.383  \\
$2^3G_3 $ 	& 3721 &  	0.383		& 3789 & 0.389 \\
$\theta_{2G}$ 	& -40.23$^\circ$ & 		& -39.93$^\circ$ & \\ 
\hline
\hline
\end{tabular}
\end{table}

\section{Radiative Transitions}
\label{sec:emtransitions}

Radiative transitions have the potential to give information that could help identify 
newly discovered states.  They are sensitive to the internal structure of states and can be particularly sensitive to ${^3}L_L - {^1}L_L$ mixing for states with $J=L$.
However, in general, charm mesons lie above the OZI decay threshold so will have
much larger strong decay partial widths than electromagnetic partial widths so in
practice we expect the usefulness of electromagnetic transitions to be limited.  Nevertheless, 
in this section we calculate E1 and M1 radiative widths.  
The partial width for an E1 radiative transition between
states in the nonrelativistic quark model is given by 
\cite{Kwo88}
\begin{widetext}
\begin{equation}
\Gamma( 
{\rm n}\, {}^{2{S}+1}{\rm L}_{J} 
\to 
{\rm n}'\, {}^{2{S}'+1}{\rm L}'_{{J}'}  
+ \gamma) 
 =  \frac{4}{3}\,  \langle e_Q \rangle^2 \, \alpha \,
k_\gamma^3 \,   
C_{fi}\,
\delta_{{S}{S}'} \delta_{{L}{L'\pm1}} \, 
|\,\langle 
{\rm n}'\, {}^{2{S}'+1}{\rm L}'_{{J}'} 
|
\; r \; 
|\, 
{\rm n}\, {}^{2{S}+1}{\rm L}_{J}  
\rangle\, |^2  
\ ,
\end{equation}
\end{widetext}
where 
\begin{equation}
\langle e_Q \rangle = {{m_c e_q - m_q e_{\bar{c}} }\over{m_q +m_c}}
\end{equation}
$e_c= 2/3$ is the $c$-quark charge and $q$ refers to the $u$, $d$ and $s$-quarks with charges
$e_u= 2/3$, $e_d=-1/3$ and $e_s=-1/3$ respectively,
 in units of $|e|$,
$\alpha$ is the fine-structure constant,
$k_\gamma$ is the photon's energy, and the angular momentum matrix element, $C_{fi}$, is given by
\begin{equation}
C_{fi}=\hbox{max}({L},\; {L}') (2{J}' + 1)
\left\{ { {{L}' \atop {J}} {{J}' \atop {L}} {{S} \atop 1}  } \right\}^2 
\end{equation}
where $\{ {\cdots \atop \cdots} \}$ is a 6-$j$ symbol.
The matrix elements 
$\langle {n'}^{2{S}'+1}{L}'_{{J}'} |\; r \; 
| n^{2{S}+1}{L}_{J}  \rangle$ 
 were evaluated 
using the wavefunctions given by the relativized quark model \cite{Godfrey:1985xj}.
Relativistic corrections are implicitly included in these E1 
transitions through Siegert's theorem \cite{Sie37,McC83,Mox83}, 
by including spin-dependent interactions in the Hamiltonian used to 
calculate the meson masses and wavefunctions.

Radiative transitions which flip spin are described by magnetic dipole
(M1) transitions.  The rates for magnetic dipole transitions between $S$-wave states in 
heavy-light bound states are given in the nonrelativistic approximation 
by \cite{JDJ,Nov78,ES}
\begin{widetext}
\begin{equation}
\Gamma(n^{2S+1}L_J  \to  {n'} ^{2S'+1}L_{J'} + \gamma)  
 =  {{\alpha}\over 3}  \; k_\gamma^3 
({2J'+1}) \delta_{S,S'\pm 1}
 \, 
 |\, \langle f | \; \frac{e_q}{m_q} j_0(k_\gamma r \frac{m_{\bar{q}}}{m_q+m_{\bar{q}} })
 -\frac{e_{\bar{q}}}{m_{\bar{q}}}   j_0(k_\gamma r \frac{m_{q}}{m_q+m_{\bar{q}} }) \; |\,  i \rangle\, |^2 
\end{equation}
\end{widetext}
where 
$e_q$, the quark charges, and $m_q$, the quark masses were given above,
$L=0$ for $S$-waves
and  $j_0(x)$ is the spherical Bessel function.
Transitions in which the principle quantum number 
changes are referred to as hindered transitions as they 
are not allowed in the non-relativistic limit due to the orthogonality 
of the wavefunctions. 
M1 transitions, especially hindered transitions,
are notorious for their sensitivity to relativistic 
corrections \cite{m1}.   
In our calculations the wavefunction orthogonality is broken 
by including a smeared  hyperfine interaction directly in the 
Hamiltonian so that the $^3S_1$ and $^1S_0$ states have slightly 
different wavefunctions.

The E1 and M1 radiative widths are given in Tables~\ref{tab:D_12S}-\ref{tab:Ds_1P} when they are large enough
that they might be observed.  More complete results are given in the supplementary material.  The tables in
the supplementary material also include the matrix elements for the benefit of the interested reader.   
The predicted masses given in 
Tables~\ref{tab:SPmasses} and \ref{tab:DFGmasses} are used for all states.  
The photon energies were calculated using the predicted masses, but assuming these masses are all slightly shifted with respect to the measured masses, the phase space should remain approximately correct.

Given the sensitivity of radiative transitions to details of the models
precise measurements of electromagnetic transition rates would 
provide stringent tests of the various calculations and predictions that have appeared in the
literature.

\section{Strong Decays}
\label{sec:strongdecays}

\begin{table*}[ht]
\caption{Light meson masses and effective harmonic oscillator parameters, $\beta_{eff}$, 
used in the calculation of strong decay widths.  The experimental values of the masses are 
taken from the Particle Data Group (PDG) \cite{Olive:2014kda}.  The input value of 
the $\pi$ mass is the weighted average of the experimental values of the $\pi^0$ 
and $\pi^\pm$ masses, and similarly for the input values of the $K$ and $K^*$ masses.  
All effective $\beta$ values are taken to be 0.4 for the light mesons.
\label{tab:lightmasses}}
\begin{ruledtabular}
\begin{tabular}{lllll}
Meson			&	State			& $M_{input}$ (MeV)	& $M_{exp}$ (MeV) \cite{Olive:2014kda}							& $\beta_{eff}$ (GeV)	\\
\hline \hline
$\pi$				& $1^1S_0$		& 138.8877			& $134.8766\pm0.0006~(\pi^0)$, $139.57018\pm0.00035~(\pi^\pm)$	& 0.4			\\
$\eta$			& $1^1S_0$		& 547.862				& $547.862\pm0.018$ 										& 0.4			\\
$\eta^{\prime}$		& $1^1S_0$		& 957.78				& $957.78\pm0.06$ 											& 0.4			\\
$\rho$			& $1^3S_1$		& 775.26				& $775.26\pm0.25$ 											& 0.4			\\
$\omega$			& $1^3S_1$		& 782.65				& $782.65\pm0.12$ 											& 0.4			\\
$\phi$			& $1^3S_1$		& 1019.461			& $1019.461\pm0.019$										& 0.4			\\
\hline
$K$				& $1^1S_0$		& 494.888				& $497.614\pm0.024~(K^0)$, $493.677\pm0.016~(K^\pm)$  			& 0.4			\\
$K^*$			& $1^3S_1$		& 894.36				& $895.81\pm0.19~(K^{*0})$, $891.66\pm0.26~(K^{*\pm})$ 			& 0.4			\\
\end{tabular}
\end{ruledtabular}
\label{tab:udsparams}
\end{table*}

For states above the $D\pi$ and $DK$ thresholds we calculate the strong decay widths 
of excited charm and charm-strange mesons using the $^3P_0$ quark pair 
creation model \cite{Micu:1968mk,Le Yaouanc:1972ae,Ackleh:1996yt,Blundell:1995ev,Barnes:2005pb}.  
There are a number of predictions for charm meson widths in the literature using 
the $^3P_0$ model \cite{Close:2005se,Zhang:2006yj,Sun:2009tg,Song:2014mha,Song:2015fha,Li:2010vx,Sun:2013qca,Yu:2014dda,Ferretti:2015rsa} 
and other models 
\cite{eichten93,DiPierro:2001uu,Zhong:2008kd,Zhong:2009sk,Colangelo:2012xi,Xiao:2014ura,Wang:2014jua,Colangelo:2006rq} 
but we believe that this work represents the most complete analysis of excited charm meson strong decays to date.
The details of the notation and conventions used in our $^3P_0$ model calculations are
given in the Appendix of Ref.~\cite{Godfrey:2015dia} to which we refer the interested reader.

We use the calculated charm and charm-strange meson masses 
listed in Tables~\ref{tab:SPmasses} and \ref{tab:DFGmasses}. For the light mesons we used the
measured masses listed in Table~\ref{tab:lightmasses}.
For the charm and charm-strange mesons 
we use harmonic oscillator (HO) wave functions with the effective harmonic oscillator parameter, 
$\beta_{eff}$, obtained 
by equating the rms radius of the harmonic oscillator wavefunction for the specified $(n,l)$ 
quantum numbers to the rms radius of the wavefunctions calculated using the relativized 
quark model of Ref.~\cite{Godfrey:1985xj}.  A previous study \cite{Godfrey:1986wj} found 
that using HO wavefunctions with the fitted oscillator parameters gave results similar 
to those calculated using the exact relativized quark model wavefunctions but 
were far more computationally efficient.  It was also found that the predictions of
the $^3P_0$ model were similar to those of the flux-tube breaking model \cite{Kokoski:1985is}
when using the same wavefunctions in both calculations \cite{Godfrey:1986wj,Blundell:1995ev,Blundell:1995au}.
The widest variation in results occurs going from using either the exact relativized quark model
wavefunctions or HO wavefunctions with $\beta_{eff}$'s to using constant $\beta$'s for
all states \cite{Godfrey:1986wj} (compare also the results from Ref.~\cite{Godfrey:2015dia}
to those of Ref.~\cite{Ferretti:2013vua}).  This is because the decay amplitudes are dominated
by the overlap of the three meson wavefunctions and the HO wavefunctions with $\beta_{eff}$'s
are a good representation of the exact wavefunctions for this purpose.
The effective harmonic oscillator wavefunction parameters, $\beta_{eff}$, 
used in our calculations
are listed in Tables~\ref{tab:SPmasses} and \ref{tab:DFGmasses}.
For the light mesons we use the universal value of $\beta_{eff}=0.4$~GeV given in Table~\ref{tab:lightmasses} (see below for 
an additional comment).
In our calculations we use the constituent quark masses 
$m_c=1.628$~GeV, $m_s=0.419$~GeV, 
and $m_q=0.220$~GeV ($q = u, d$).  Finally, we use ``relativistic phase space'' as 
described in Ref.~\cite{Blundell:1995ev,Ackleh:1996yt} and in the Appendix
of Ref.~\cite{Godfrey:2015dia}.

Typical values of the parameters $\beta_{eff}$ and $\gamma$, the quark pair creation
amplitude of the $^3P_0$ model,  are found from fits to light meson
 decays \cite{Close:2005se,Blundell:1995ev}. The predicted widths are fairly 
 insensitive to the precise values used for $\beta_{eff}$ provided $\gamma$ is appropriately rescaled.  
 However $\gamma$ can vary as much as 30\% and still give reasonable overall fits of light meson 
 decay widths \cite{Close:2005se}. This can result in factor of two changes to 
 predicted widths, both smaller or larger.  In our calculations of $D_s$ meson strong decay 
 widths in \cite{Godfrey:2014fga}, we used a value of $\gamma=0.4$, which has also been found 
 to give a good description of strong decays of charmonium \cite{Barnes:2005pb,Close:2005se}.  
   This scaling of the value of $\gamma$ in different meson sectors has been studied 
  in \cite{Segovia:2012}.   
  The resulting strong decay widths are listed in 
  Tables~\ref{tab:D_12S}-\ref{tab:Ds_2G3} although we only show decays that have branching ratios greater than
  $\sim$~1\% or decays to simple final states such as $D\pi$ or $D K$ that might be
  easier to observe.  More complete tables of results are given in the supplementary material.

\begin{table}
\caption{Partial widths and branching ratios
for strong and electromagnetic decays for the  $1S$, $2S$ and $3S$ charm mesons. The initial state's mass is given 
in GeV and is listed below the state's name in column 1.  
We only show radiative transitions that are likely to be observed 
and likewise generally
do not show strong decay modes which have BR $\lesssim$ 1\%  although they are included
in calculating the total width and are included in the supplementary material. The matrix
elements for radiative transitions are given in the supplementary material. 
Details of the calculations are given in the text.  
\label{tab:D_12S}}
\begin{ruledtabular}
\begin{tabular}{l l  c c } 
Initial 	& Final  	& Width ($cu$, $cd$) &  B.R. ($cu$, $cd$)	\\
state 		& state 	& (MeV) 	& (\%)		\\
\hline \hline
$D^* $ 		& $D \gamma$				& 0.106,	0.0108	& 45.9, 8.0		 \\
2041			& $D\pi$ 	 			& 0.125	& 54.1, 92.1	 \\
			&	Total					& 0.231, 0.136 & 100		 \\
\hline
$D(2^3S_1) $ 	& $D \gamma$ & 0.60, 0.10	& 0.6, 0.1	 \\
2643			& $D\pi$ 	& 26.1	& 25.6	 \\
			& $D\eta$ 		& 3.85	& 3.8	 \\
\end{tabular}
\end{ruledtabular}
\end{table}

\begin{table}
\caption{Partial widths and branching ratios
for strong and electromagnetic decays for the $4S$ charm mesons. See the caption to
Table~\ref{tab:D_12S} for further explanations.
\label{tab:D_4S1}}
\begin{ruledtabular}
%

\section{Classification of the Observed Charm Mesons}
\label{sec:charm_classification}

The Belle, BaBar and LHCb collaborations have increased our knowledge of charm mesons
considerably in recent years which has spawned a large number of theory papers attempting 
to categorize these new states.  We list these new states and their properties 
in Table~\ref{tab:obs_charm}.

\begin{table*}[t]
\caption{Recently observed charm mesons.  The first uncertainty is statistical, the second 
uncertainty is due to systematic uncertainties and the third when included is for model dependent 
uncertainties.
\label{tab:obs_charm}}
\begin{tabular}{llllll} \hline \hline
State 			&  $J^P$	& Observed Decays & Mass (MeV) 	& Width (MeV) 	  & References  \\ \hline
$D_J(2550)^0$	& $0^-$	& $D^{*+}\pi^-$	& $2539.4\pm 4.5\pm 6.8$	& $130\pm 12 \pm 13$ & BaBar \cite{delAmoSanchez:2010vq} \\
$D_J(2580)^0$ 	& 		& $D^{*+}\pi^-$ & $2579.5\pm 3.4 \pm 3.5 $   & $177.5 \pm 17.8 \pm 46.0 $ 	&   LHCb \cite{Aaij:2013sza}\\ 
$D_J^*(2600)^0$	&		& $D^+\pi^-$ & $ 2608.7\pm 2.4 \pm 2.5 $ & $ 93\pm 6 \pm 13$ & BaBar \cite{delAmoSanchez:2010vq} \\
				&		& 			&		& $\Gamma (\to D^+\pi^-)/\Gamma(\to D^{*+} \pi^-) = 0.32 \pm 0.02 \pm 0.09$ &  BaBar \cite{delAmoSanchez:2010vq}  \\
$D_J^*(2650)^0$ & 		& $D^{*+}\pi^-$ & $2649.2 \pm 3.5 \pm 3.5$    & $140.2 \pm 17.1 \pm 18.6  $ &  LHCb \cite{Aaij:2013sza} \\
$D_J(2750)^0$		&		& $D^{*+}\pi^-$ & $ 2752.4 \pm 1.7 \pm 2.7$  & $71 \pm 6 \pm 11$  & BaBar \cite{delAmoSanchez:2010vq} \\
$D_J(2740)^0$ 	&  		& $D^{*+}\pi^-$ & $2737.0\pm 3.5 \pm 11.2 $    & $73.2 \pm 13.4 \pm 25.0 $ 	& LHCb \cite{Aaij:2013sza}  \\
$D_J^*(2760)^0$	&		& $D^{*+}\pi^-$ & $2761.1\pm 5.1 \pm 6.5$	& $ 74.4 \pm 3.4 \pm 37.0$ & LHCb \cite{Aaij:2013sza}\\
				&		& $D^{+}\pi^-$ & $2760.1\pm 1.1 \pm 3.7$	& $ 74.4 \pm 3.4 \pm 19.1$ & LHCb \cite{Aaij:2013sza}\\
				&		& $D^+\pi^-$	& $2763.3 \pm 2.3 \pm 2.3 $  & $ 60.9 \pm 5.1 \pm 3.6$ & BaBar \cite{delAmoSanchez:2010vq} \\
				&		& 	&  & $\Gamma (\to D^+\pi^-)/\Gamma(\to D^{*+} \pi^-) = 0.42 \pm 0.05 \pm 0.11$ &  BaBar \cite{delAmoSanchez:2010vq}  \\
$D_J^*(2760)^+$ & 		& $D^0\pi^+$  & $2771.7 \pm 1.7 \pm 3.8  $   & $66.7  \pm 6.6 \pm 10.5  $ 	&  LHCb \cite{Aaij:2013sza} \\
$D_1^*(2760)^0$	& $1^-$	& $D^+\pi^-$	& $2781\pm 18 \pm 11\pm 6$ & $177 \pm 32\pm 20\pm 7$ & LHCb \cite{Aaij:2015vea} \\
$D_3^*(2760)^-$	& $3^-$	& $\bar{D}^0\pi^-$	& $2798\pm 7 \pm 1\pm 7$ & $105 \pm 18\pm 6\pm 23$ & LHCb \cite{Aaij:2015sqa}\footnotemark[1] \\
$D_J(3000)^0$ & 		& $D^{*+}\pi^-$  	& $2971.8 \pm 8.7$  		&  $188.1 \pm 44.8$ &	LHCb \cite{Aaij:2013sza}\\
$D_J^*(3000)^0$ 	& 		& $D^{+}\pi^-$ & $3008.1 \pm 4.0$   		&  $110.5 \pm 11.5$ &	LHCb \cite{Aaij:2013sza}\\
\hline
\hline
\end{tabular}
\footnotetext[1]{We quote the results from the isobar analysis.}
\end{table*}

The $1P$ multiplet is well established and its measured properties agree 
well with theoretical expectations.
Theory expects in the heavy quark limit two doublets, the $j=1/2$ doublet composed of a $0^+$ and $1^+$ state,
corresponding to the $D_0(2400)$ and $D_1(2430)$ states, that decay via $S$-wave and are broad,
and the $j=3/2$ doublet composed of a $1^+$ and $2^+$ state,
corresponding to the $D_1(2420)$ and the $D_2^*(2460)$
 that decay via $D$-wave and are relatively narrow.  These states have been discussed
 in the literature (see for example \cite{Godfrey:2005ww}).
 
It is the large number of newly observed states that concerns us here.  
Given the success of quark model calculations in describing the $1P$ states we use the 
QM predictions of the previous sections to classify these new states.  The success of these efforts can be used
to gauge the reliability of our predictions for the properties of undiscovered states. 
These states have been discussed in numerous papers 
\cite{Close:2006gr,Zhang:2006yj,Colangelo:2006rq,Zhong:2009sk,Song:2015fha,Chen:2009zt,Li:2009qu,Zhang:2009nu,Li:2010vx,Wang:2010ydc,Wang:2013tka,Wang:2013mml,Lu:2014zua,Sun:2013qca,Chen:2015lpa,Yu:2014dda,Badalian:2011tb,vanBeveren:2006st,vanBeveren:2009jq,Xiao:2014ura,Zhong:2010vq}. 

\subsection{The $D_J(2550)^0$ and $D_J^*(2600)^0$ States}

Both the BaBar \cite{delAmoSanchez:2010vq} and LHCb \cite{Aaij:2013sza} collaborations 
have reported states around 2550~MeV and 2600~MeV whose
measured properties we list in Table~\ref{tab:obs_charm}.  The only states expected to fall
in this mass region are the $2^1S_0(c\bar{q})$ and $2^3S_1(c\bar{q})$ states. 

Starting with the $D_J(2550)^0$,
the masses reported by the two experiments are inconsistent at the $2\sigma$ level.  
However BaBar and LHCb measure the widths to be 130~MeV and 178~MeV with large errors.  Given
the large measured widths we believe that the experiments are seeing the same state but that it is
difficult to extract masses precisely. With this assumption we average the masses
and widths to obtain $M(D_J(2550)^0)=2559$~MeV and $\Gamma(D_J(2550)^0)=154$~MeV but don't 
attempt to estimate uncertainties given the naivety of the averaging. In addition LHCb makes the point 
that extracting its parameters is complicated \cite{Aaij:2013sza}. 
 Both experiments identify the  $D_J(2550)^0$ as the $2^1S_0(c\bar{q})$ state.  We predict
 the mass and width of the $2^1S_0(c\bar{q})$ state to be 2581~MeV and 80~MeV respectively.  
 We note that we predict the $1^3S_1$ mass to be 2041~MeV which is $\sim 34$~MeV greater 
 than its observed mass so that if we rescale the $2^1S_0(c\bar{q})$ mass by this amount we
 obtain 2547~MeV which is consistent with the observed average value.  The calculated width
 is significantly smaller than the average of the measured widths.  However, due to both
 the  uncertainty in the theoretical width predictions and the 
 large uncertainty in the measured width we consider the agreement between theory and experiment
 to be acceptable and conclude that the $D_J(2550)^0$ is the $2^1S_0(c\bar{q})$ state.
 
 We follow the same logic in  comparing the measured properties of the $D_J^*(2600)^0$ to the predicted 
 properties for the $2^3S_1(c\bar{q})$.  Again the masses reported by BaBar and LHCb for the 
  $D_J^*(2600)^0$ are incompatible at the level of $2\sigma$ and the measured widths 
  are 93 and 140~MeV respectively with large errors.  Averaging the  masses and
  widths measured by the two experiments we obtain $M(D_J^*(2600)^0)=2629$~MeV and 
  $\Gamma(D_J^*(2600)^0)=117$~MeV and again we do not assign errors given the lack of justification
  for using simple averages.  We predict
 the mass and width of the $2^3S_1(c\bar{q})$ state to be 2643~MeV and 102~MeV respectively. 
Again,  if we rescale the $2^3S_1(c\bar{q})$ mass down by 34~MeV we obtain 2609~MeV.  Further, we find that 
$\Gamma (D^*(2S)\to D\pi)/\Gamma(D^*(2S)\to D^{*} \pi) = 0.44$ versus 
the BaBar \cite{delAmoSanchez:2010vq}
measurement of $0.32 \pm 0.02 \pm 0.09$.  We conclude
that the $D_J^*(2600)^0$ properties are consistent with the predicted properties for the $2^3S_1(c\bar{q})$.

To summarize,  our results support the identification of the $D_J(2550)^0$ and $D_J^*(2600)^0$
with the $2^1S_0(c\bar{q})$ and $2^3S_1(c\bar{q})$ respectively made by the
BaBar \cite{delAmoSanchez:2010vq} and LHCb \cite{Aaij:2013sza} collaborations and 
previous theoretical studies 
\cite{Li:2010vx,Wang:2010ydc,Chen:2011rr,Wang:2013tka,Lu:2014zua,Yu:2014dda,Zhong:2010vq,Chen:2015lpa}.

\subsection{The $D_J(2750)^0$, $D_1^*(2760)^0$ and $D_3^*(2760)^0$  States}

The BaBar \cite{delAmoSanchez:2010vq} and LHCb \cite{Aaij:2015vea,Aaij:2015sqa,Aaij:2013sza} 
collaborations have observed a number of new states in the mass region
of 2740 to 2800~MeV.  They can be grouped into the un-natural parity 
$D_J(2750)$ seen in the $D^*\pi$ final 
state and some number of natural parity states collectively labelled the $D_J^*(2760)$ which has recently
been resolved by LHCb into two states with $J^P=1^-$ and $3^-$ labelled the 
$D_1^*(2760)$ \cite{Aaij:2015vea} 
and $D_3^*(2760)$ \cite{Aaij:2015sqa} seen in the $D\pi$ final state.  The states predicted
to be closest in mass to these states are the $1D$ states.  The next nearest states are the 
$2P$ multiplet which is expected to lie in the 2900 to 2950~MeV mass region but whose quantum
numbers are inconsistent with the recent LHCb measurements \cite{Aaij:2015vea,Aaij:2015sqa}.
We will therefore concentrate on the expected properties of the $1D$ multiplet.  As we
did for the $2S$ multiplet, for the purposes of comparing our mass predictions to the
measured masses, we will rescale our predictions down by the difference between predicted and 
measured masses for the $D^*(1S)$ state of 34~MeV.

Starting with the $D_3^*(2760)^0$ state, the measured mass and width from the LHCb isobar analysis
are $2798 \pm 10$~MeV and $105\pm 30$~MeV  \cite{Aaij:2015sqa} respectively
where for simplicity we have combined the statistical,
experimental and systematic errors in quadrature. These values should be compared 
to the predicted mass (after rescaling) and width of the 
$1^3D_3(c\bar{q})$ state which are 2799~MeV and 51~MeV respectively.  
The agreement between experiment and our predictions for both the mass and total width are
satisfactory so we identify the $D_3^*(2760)^0$ as the $1^3D_3(c\bar{q})$ state.

Similarly we compare the measured mass and width for the $D_1^*(2760)^0$ which
are $M=2781 \pm 22$~MeV and $\Gamma=177 \pm 38$~MeV \cite{Aaij:2015vea} where again we have 
combined the statistical, experimental and systematic errors in quadrature.  The predicted
mass (after rescaling) and width are 2774~MeV and 233~MeV respectively which are in good
agreement with the measured properties of the $D_1^*(2760)^0$.  We therefore identify
the $D_1^*(2760)^0$ as the $1^3D_1(c\bar{q})$ state.

The final state in this mass region is the un-natural parity $D_J(2750)^0$ state.  For the purposes of this
discussion we average the BaBar \cite{delAmoSanchez:2010vq} and LHCb \cite{Aaij:2013sza}
measurements to obtain $M=2744.7$~MeV and $\Gamma=72.1$~MeV.  The measured mass is marginally
inconsistent with the predicted masses of the two $J=2$ $D$-wave states;
$M(1D_2)=2782$ and $M(1D'_2)=2811$~MeV
(after rescaling).  The predicted widths are $\Gamma(1D_2)=105$~MeV and $\Gamma(1D'_2)=244$  respectively.
Considering both the experimental uncertainty, which  at least for the LHCb measurement 
is large, and the theoretical uncertainty, it is reasonable to identify the 
the $D_J(2750)^0$ with the $1D_2(c\bar{q})$ state.   This identification
can be verified by observing the  $D_J(2750)^0$ in other decay modes.  For example,
 $BR (1D_2\to D\rho)=58$\% while  $BR (1D_2'\to D\rho)=2.4$\%.  This should be a relatively
 simple state to observe.  The second largest decay mode for the $1D_2'$ is
 $BR (1D_2' \to D^*\pi)\sim38$\%
 vs $BR( 1D_2 \to D^*\pi)\sim 20$\%.    
 Finally,  the next largest BR for $1D_2'$ is $BR (1D_2'\to D(1^3P_2)\pi)=37$\%
 vs  $BR (1D_2\to D(1^3P_2)\pi)< 1$\%, which is another discriminator between these two possibilities
although this final state  is likely to be difficult to observe. 

We conclude that  the  $D_1^*(2760)^0$,  $D_3^*(2760)^0$ and
$D_J(2750)^0$ states can be identified with the $1^3D_1(c\bar{q})$, $1^3D_3(c\bar{q})$
and $1D_2(c\bar{q})$ quark model states, respectively.  Our identification of the $D_J(2750)^0$
with the $1D_2(c\bar{q})$  is consistent with other studies 
\cite{Lu:2014zua,Zhong:2010vq,Wang:2010ydc,Xiao:2014ura,Song:2015fha} 
although in some cases they label this state with the prime.

With three of the four $1D$ states observed by the BaBar and 
LHCb collaborations there is one remaining $1D$ state to be found.  However, with four overlapping states
it is not an easy task to disentangle them based solely on mass and total width measurements
and to make precise measurements of their properties. Measuring BR's will be useful for
solidifying the spectroscopic assignments given above and resolving ambiguities and
inconsistencies.   For example,  the LHCb collaboration resolved the $D^*_J(2760)$ into
$J=1$ and $J=3$ states.  Our calculations predict $\Gamma(D\pi)/\Gamma(D^*\pi)=1.8$ 
for the $1^3D_1(c\bar{q})$ and 1.3 for $1^3D_3(c\bar{q})$.  These are inconsistent with 
the BaBar measurement of $0.42 \pm 0.05 \pm 0.11$ for the $D_J^*(2760)^0$ \cite{delAmoSanchez:2010vq}.
It is possible that a new measurement will resolve this discrepancy but it is also possible
that the $D^*\pi$ signal contains significant contributions from broad overlapping $D_2$
states.  Further useful information about these states can be obtained by measuring
BR's into other final states such as to $D\rho$ and $D_sK$ with the relevant BR's given 
in Table~\ref{tab:D_1D1}.

\subsection{The $D_J(3000)^0$  and $D_J^*(3000)^0$  States}

LHCb has reported two states around 3000~MeV, the natural parity state
$D_J^*(3000)^0$ with mass = $3008.1 \pm 4.0$~MeV  and width = $110.5 \pm 11.5$~MeV
and the un-natural parity
state $D_J(3000)^0$ with mass = $2971.8 \pm 8.7$~MeV  and width = $188.1 \pm 44.8$~MeV \cite{Aaij:2013sza}.
Our calculations expect the $2P$, $3S$ and $1F$ multiplets to lie in this mass region
consisting of the natural parity $2^3P_2$, $2^3P_0$, $3^3S_1$, $1^3F_4$ and $1^3F_2$ states
and the un-natural parity $2P_1'$, $2P_1$, $3^1S_0$, $1F_3'$ and $1F_3$ states.  
All of these states are expected to have widths in the range of 114 to 270~MeV which, 
given the theoretical and experimental uncertainties, cannot by themselves be used to rule
out any of the possibilities.  To help us narrow the possibilities we summarize the predicted
properties of the $2P$, $3S$, and $1F$ multiplets in Table~\ref{tab:D3000}.  We list
BR's for the simpler final states under the assumption that they will be easiest to observe. 
We also include some final states with larger BR's as they contribute to the total width but expect
that in many cases they will be challenging to reconstruct.

\begin{table*}[ht]
\caption{Properties of the  $2P$, $3S$, and $1F$ charm meson multiplets.  The predicted masses
listed here have been shifted down by 34~MeV, the difference between the predicted and measured $D^*$ masses.
We list BR's of the simplest final states and in some cases final states with the largest BR's.  
Blank entries represent either forbidden decays or BR's too small to be included.
\label{tab:D3000}}
\begin{tabular}{lccrrrrrrrrrr} \hline \hline
State 	& Mass (MeV) & Width (MeV) &  \multicolumn{10}{c}{Branching Ratios  (\%)}  \\
	&	&	 & $D\pi$  & $D^*\pi$ & $D\rho$  & $D^*\rho$ & $D\omega$ & $D^*\omega$ & $D_sK$ & $D_s^*K$ & $D_sK^*$ & $D_s^*K^*$  \\ \hline
 \multicolumn{13}{c}{Natural parity states} \\ 
$2^3P_2$ & 2923	& 114 & 4.4 & 15 & 9.3 & 23   & 3.0 & 8.1 & 3.2 & 4.9 & 0.3 &    \\
$2^3P_0$ & 2897	& 190 & 13.4 &   &      & 16.8 &  & 5.4 & 0.4 &  &  &  	\\
$3^3S_1$ & 3076	& 103 & 3.1 & 5.4 & 0.8 & 3.2 & 0.2 & 1.3 &  & 0.6 & 0.9 & 6.3	\\
$1^3F_4$ & 3079	& 129 & 12.2 & 11.8 & 3.1 & 45.7 & 1.0 & 14.8 & 0.8 & 0.6 & & 1.2\\
$1^3F_2$ & 2098 & 243 & 9.5 & 7.6 & 6.7 & 6.6 & 2.2 & 2.1 & 3.3 & 2.1 & 0.8 & 0.2  \\ 
 \multicolumn{13}{c}{Un-natural parity states} \\
$2P_1$ 	& 2890	& 125 & 	& 30.3 & 2.7 & 19.3 & 0.9 & 6.5 &  & 7.2 & 11.4 &    \\
$2P_1'$ & 2924	& 212 & 	& 10.2 & 8.9 & 11 &  & 3.5 &  & 2.1 & 1.9 &    \\
$3^1S_0$ & 3034 & 106 & 	& 4.9 & 	& 9.3 & & 3.3 &  & 2.3 & 2.2 & 3.2  \\
$1F_3'$ & 3109 	& 269 & 	& 17.2 & 3.6 & 11.0 & 1.2 & 3.6 &  & 5.2 &  & 0.4    \\
$1F_3$ &  3074	& 126 & 	& 20 & 31 & 20.3 & 10.2 & 6.6 &  & 0.9 & 3.3 & 0.4 \\
\hline
\hline
\end{tabular}
\end{table*}

Before proceeding we note that LHCb comments that the resonance parameters are strongly 
correlated to the background parametrization and they don't include the broad 
$D_0^*(2420)$ \cite{Aaij:2013sza}.  
Thus, one should be cautious in how literally one takes the LHCb values.

Given the general uncertainties of our predictions it is difficult to  make definitive 
spectroscopic assignments for the $D_J^*(3000)$ and $D_J(3000)$ states.  At best we can narrow 
down the possibilities, present our most likely assignment and suggest future measurements that could 
uniquely identify these states.  With this caveat we note that in general our mass predictions
tend to overestimate masses of excited states rather than underestimate them. The $2P$ multiplet
lies around 2900~MeV, 100~MeV below the observed masses so we consider it less likely that
the  $D_J^*(3000)$ and $D_J(3000)$ are $2P$ states.  

For the natural parity states, this leaves the $3^3S_1$, $1^3F_4$ and $1^3F_2$.  We calculate
a total width for the $1^3F_2$ of 243~MeV versus the measured width of $110.5 \pm 11.5$~MeV.  
As we have stated previously,  we would not be 
surprised if our width predictions are off by a factor of two.  Nevertheless it is likely
that the properties of the $1^3F_2$ are inconsistent with those of the  $D_J^*(3000)$.  
As a final discriminator we consider signal strengths assuming that the state with the
largest expected signal strength is the state most likely to first be observed.  
Signal strengths are a product of the production cross section and final state BR.  
We surmise that the cross section for orbitally excited states are suppressed compared to 
states with small orbital angular momentum but we don't know how to accurately 
calculate the production cross section for charm mesons so only consider the final state
BR.  The BR's for the two remaining possibilities are  
$BR(D^*_1(3^3S_1)  \to D\pi ) \simeq 3 \%$ vs $BR(D^*_4(1^3F_4) \to D\pi) \simeq 12 \% $.  
On this basis we tentatively identify the $D_J^*(3000)$ as the $D^*_4(1^3F_4)$ state but 
note that this conclusion is based on a number of unsubstantiated assumptions.  

The key to confirming this identification will be measuring BR's to other final states
and ratios of BR's.  Observing the $D_J^*(3000)$ in the $D^*\pi$ final state would rule out 
the $2^3P_0$.  Measuring the ratio $R=BR(D^*\pi)/BR(D\pi)$ could narrow down the options.
Large ratios would imply the $2^3P_2$ with $R\sim 3.4$ or $3^3S_1$ with $R\sim 1.7$ while 
a small ratio would imply $1^3F_2$ with $R\sim 0.8$ and a ratio $\sim 1$ would imply the 
$1^3F_4$.  Undoubtably the experimental errors will be large to start with and therefore not
precise enough, also given the theoretical uncertainties, to narrow down the possibilities 
to one specific state.  For example, the $1^3F_4$ decays almost half the time to $D^*\rho$
while the $3^3S_1$ and $1^3F_2$ have much smaller BR's to this final state.  Finally,
the $1^3F_2$ has a much larger BR to $D\rho$ than does the $3^3S_1$.  Thus, observing 
more final states can either confirm the hypothesis that the $D_J^*(3000)$ is the $D^*_4(1^3F_4)$
or direct us to an alternative identification.

We follow the same approach when trying to identify the un-natural parity $D_J(3000)$ state. 
We will set aside the $2P_1$ and $2P_1'$ as they are likely to be too low in mass.  None of the 
three remaining states can be ruled out based on their total widths, considering
both the experimental and theoretical uncertainties.  The $3^1S_0$ mass is closest to 
the measured mass while the $1F_3$ and $1F_3'$ have much larger BR's to the observed $D^*\pi$ 
final state.  We slightly favour the $D(3^1S_0)$ identification but more information is 
needed to make a more informed identification.  For example,  if the $D_J(3000)$ were observed
in the $D\rho$ final state the $3^1S_0$ would be ruled out. 
Measuring the ratios of BR's of $R=BR(D\rho)/BR(D^*\pi)$ would provide a powerful 
discriminator between the remaining options: $2P_1$, $2P_1'$, $1F_3$ and $1F_3'$.  Numerous other
final states could be used to help identify the quark model assignment of the $D_J(3000)$
but are typically more difficult to reconstruct so are unlikely to provide useful input
in the near future.

We therefore tentatively identify the $D_J^*(3000)$ as the $D_4^*(1^3F_4)$ state and favour
the $D_J(3000)$ as the $D(3^1S_0)$ state although we do not rule out the $1F_3$ and $1F_3'$
assignments. Previous studies have come to the same conclusions \cite{Lu:2014zua}.  However, 
Ref.~\cite{Xiao:2014ura} agrees with our identification of the $D_J^*(3000)$ but suggests that
the $D_J(3000)$ is a $2P_1$ state, Ref.~\cite{Yu:2014dda} identifies the 
$D_J^*(3000)$ as either the $D_4^*(1^3F_4)$ or $D_2^*(1^3F_2)$ and the $D_J(3000)$ as either
the $D_3(1F_3)$ or $D_1^\prime(2P_1^\prime)$,    
and Ref.\cite{Sun:2013qca} argues that the $D_J(3000)$ and
$D_J^*(3000)$ are the $2P_1$ and $2^3P_0$ respectively. Still other assignments appear 
in the literature \cite{Song:2015fha}.  Clearly further measurements of
other BR's will be needed to settle the issue.

\section{Classification of the Observed Charm-Strange Mesons}
\label{sec:charm_strange_classification}

We summarize the properties of the recently observed charm-strange mesons in Table~\ref{tab:obs_charm_strange}.

\begin{table*}[t]
\caption{The recently observed charm-strange mesons.
The first uncertainty is statistical, the second 
uncertainty is due to experimental systematic effects and the third, when given,
is due to model variations.
\label{tab:obs_charm_strange}}
\begin{tabular}{llllll} \hline \hline
State 			& $J^P$ & Observed Decays & Mass (MeV)  & Width (MeV)  & References  \\ \hline
$D_{s1}^*(2700)^+$	& $1^-$	& $D^0 K^+$	& $2699^{+14}_{-7}$ & $127^{+24}_{-19}$ & BaBar \cite{Lees:2014abp} \\
					&		&	&  & ${{\Gamma(D^*K)}\over{\Gamma(DK)}}=0.91 \pm 0.13 \pm 0.12$ & BaBar \cite{Aubert:2009ah}   \\
$D_{s1}^*(2700)^+$	& $1^-$		& $D^+K^0$ and $D^0 K^+$	& $2709.2 \pm 1.9 \pm 4.5$ & $115.8\pm 7.3 \pm 12.1$ & LHCb  \cite{Aaij:2012pc} \\
$D_{sJ}^{*}(2860)^+$ & 	& $DK$ and $D^*K$   & $2863.2^{+4.0}_{-2.6} $  & $58\pm11$ &  PDG \cite{Olive:2014kda}\\
$D_{sJ}^{*}(2860)^+$ & 		& $DK$ and $D^*K$ & $2862 \pm 2 ^{+5}_{-2} $  & $48\pm 3 \pm 6$ &  BaBar  \cite{Aubert:2009ah}\\
					&			&  & 			& ${{\Gamma(D^*K)}\over{\Gamma(DK)}}=1.10 \pm 0.15 \pm 0.19$ &  BaBar \cite{Aubert:2009ah}\\
$D_{s1}^{*}(2860)^-$ & $1^-$	& $\bar{D}^0 K^-$  &  $2859 \pm 12 \pm 6 \pm 23$   & $159 \pm 23 \pm 27 \pm 72$   &  LHCb \cite{Aaij:2014xza,Aaij:2014baa} \\
$D_{s3}^{*}(2860)^-$ & $3^-$	& $\bar{D}^0 K^-$  &  $2860.5 \pm 2.6 \pm 2.5 \pm 6.0$   & $53 \pm 7 \pm 4 \pm 6$ &  LHCb \cite{Aaij:2014xza,Aaij:2014baa} \\
$D_{sJ}(3044)^+$ & 	& $D^*K$ &  $3044\pm 8^{+30}_{-8}$   & $239 \pm 35 ^{+46}_{-42}$  	&  BaBar \cite{Aubert:2009ah} \\
\hline
\hline
\end{tabular}
\end{table*}

\subsection{The  $D_{s1}^*(2709)^\pm$, $D_{s1}^{*}(2860)^-$ and  $D_{s3}^{*}(2860)^-$ States}

The $D_{s1}^*(2709)^\pm$  \cite{Aubert:2006mh,Brodzicka:2007aa,Aubert:2009ah,Aaij:2012pc} 
and $D_{sJ}^{*}(2860)^-$ \cite{Aubert:2006mh,Aubert:2009ah,Aaij:2012pc} 
were first 
observed by the Belle \cite{Brodzicka:2007aa}, BaBar \cite{Aubert:2006mh,Aubert:2009ah} 
and LHCb \cite{Aaij:2012pc} collaborations.  
More recently the LHCb collaboration has measured the properties of
the $D_{sJ}^{*}(2860)^-$ more precisely and found that it is comprised of two overlapping
states, the $D_{s1}^{*}(2860)^-$ and $D_{s3}^{*}(2860)^-$ with $J^P= 1^-$ and $3^-$ 
respectively \cite{Aaij:2014xza,Aaij:2014baa}.  
With this new information it was argued that the $D_{s1}^*(2709)^\pm$ is the $2^3S_1(c\bar{s})$ state
and the $D_{s1}^{*}(2860)^-$ and $D_{s3}^{*}(2860)^-$ are the $1^3D_1(s\bar{c})$ and 
$1^3D_3(s\bar{c})$ states respectively \cite{Aaij:2014xza,Aaij:2014baa,Godfrey:2014fga,Song:2014mha,Wang:2014jua}.  
The largest overall
discrepancy between theory and experiment is for the ratio 
$\Gamma(2^3S_1 \to D^*K)/\Gamma(2^3S_1 \to DK)$.  However, it was argued that this discrepancy
could be explained by treating the $D_{s1}^*(2709)^\pm$ as a mixture of $2^3S_1(c\bar{s})$ and $1^3D_1(c\bar{s})$
\cite{Godfrey:2014fga,Godfrey:2013aaa,Close:2006gr,Zhong:2009sk,Chen:2011rr,Li:2010vx,Li:2009qu,Yuan:2012ej,Wang:2013mml}
with a relatively small $2^3S_1 - 1^3D_1$ mixing angle of $\sim 10^\circ$ \cite{Godfrey:2014fga}. 
A consequence of this
identification is that there are another three excited $D_s$ states in this mass region to be 
found; the spin-singlet partner of the $D_{s1}^*(2709)^\pm$ is expected
to lie $\sim 60$~MeV lower in mass with $M(2^1S_0(c\bar{s}))\sim 2650$~MeV, a width
of $\sim 78$~MeV and decaying to $D^*K$ \cite{Godfrey:2014fga}.  
The $J=2$ states using the predicted $1D$ mass
splittings relative to the $1^3D_3$ mass give $M(D_2')\sim 2872$~MeV and $M(D_2)\sim 2846$~MeV with the 
partial widths 
given in Table~\ref{tab:d2} which were taken from Ref.~\cite{Godfrey:2014fga}.
We note that in the HQL we expect one of the $J=2$ states to be degenerate with the $1^3D_3$ 
and relatively narrow while the other $J=2$ state is expected to be degenerate with the
$1^3D_1$ and relatively broad which is consistent with our results. It will be 
interesting to see what experiment has to say about these states.

\begin{table}[t]
\caption{Partial widths for the  $1D_2$ and $1D_2^\prime$ $c\bar{s}$ mesons calculated using the 
$^3P_0$ quark pair creation model from Ref.~\cite{Godfrey:2014fga}.
The $1D_2$ and $1D_2^\prime$ masses listed here and used to calculate the partial widths
 were obtained by subtracting the predicted splittings 
from the measured  $1^3D_3$ mass.
\label{tab:d2}}
\begin{tabular}{llc} \hline \hline
State & Property   & {Predicted} \\
	  &				& (MeV)  \\
\hline 
$D_s(D_{2}')$ 		& Mass 						 	& 2872	\\	
				& ${D_s}_{2}'\to D^*K$ 				& 159			\\	
				& ${D_s}_{2}'\to DK^*$ 				& 4.4			\\	
				& ${D_s}_{2}'\to D^*K^*$ 			& 0			\\	
				& ${D_s}_{2}'\to D_s^*\eta$ 		& 21			\\	
  				& $\Gamma_{\hbox{Total}}$ 			& 184		\\	
\hline
$D_s(D_{2})$ 		& Mass 							& 2846	\\	
				& ${D_s}_{2}\to D^*K$ 				& 16		\\	
				& ${D_s}_{2}\to DK^*$ 				& 58			\\	
				& ${D_s}_{2}\to D^*K^*$ 			& 0			\\	
				& ${D_s}_{2}\to D_s^*\eta$ 			& 0.4			\\	
   				& $\Gamma_{\hbox{Total}}$ 			& 75		\\	
\hline
\hline
\end{tabular}

\end{table}

\subsection{The $D_{sJ}(3044)^\pm$ State}

The remaining new state is the $D_{sJ}(3044)^\pm$.  
This state has been studied by a number of authors 
\cite{Chen:2009zt,Sun:2009tg,Zhong:2009sk,Colangelo:2010te,Li:2010vx,Badalian:2011tb}.  
We start by noting that it has
only been seen by one experiment, BaBar \cite{Aubert:2009ah}, albeit with 6.0 standard deviation 
statistical significance. It has only been seen in the $D^*K$ final state implying it is
of unnatural parity; $0^-$, $1^+$, $2^-$, $3^+$, $4^-$, etc.  
Of all the states with these quantum numbers the predicted masses for the
$2P_1^\prime$ and $2P_1$ states are closest to the observed mass,  3038~MeV and 3018~MeV
respectively.  The $2P_1$ is expected to have a total width of 143~MeV and a BR to the
$D^*K$ final state of 43\% while the $2P_1'$ is expected to have a total width of
148~MeV with BR to $D^*K$ of 25\%.  The experimental error on the width is quite large, 
$\sim\pm 55$~MeV and there is considerable theoretical uncertainty on the predicted width 
which could  be up to a factor of 2.  As a consequence, the $D_{sJ}(3044)^\pm$ could be
either state.  Referring to Table~\ref{tab:Ds_2P1} the $DK^*$ final state
might be a useful discriminator between 
these two possibilities; the $2P_1'$ is predicted to have a BR of 22\% to $DK^*$ while for the
$2P_1$ it is predicted to be 5\%.  Another possibility is that because the states are relatively 
close together and broad, perhaps BaBar observed two overlapping states. A means of discriminating
between these possibilities is to measure BR's to different final states.  For example,
we estimate that the branching ratios of the $D_{s1}(2P_1')$ to $DK^*$ and $D_s\phi$ are
22\% and 2.8\% respectively versus 43\% and 11.3\% respectively for the  $D_{s1}(2P_1)$.
Measurement of BR's to these final states would provide a good discriminator for these 
possibilities (see also Ref.~ \cite{Sun:2009tg,Zhong:2009sk,Colangelo:2010te,Li:2010vx}).
 A final possibility is
that the observed state does decay to $DK$ but that it simply was not observed.  The
$D_{s2}^* (2^3P_2)$ state is predicted to have a mass $\sim 3048$~MeV, total width 132~MeV
with BR's to $D^*K$ and $DK$ of 23\% and 6.5\% respectively.  It may have been that the 
signal in the $DK$ final state was simply too small to see with limited statistics.  

While we consider it most likely that the $D_{sJ}(3044)^\pm$ is a member
of the $2P(c\bar{s})$ multiplet we mention other possibilities for completeness.  Other
unnatural parity states with masses not too far from the $D_{sJ}(3044)^\pm$ are
the $1F_3$ with $M=3186$~MeV and $\Gamma=183$~MeV, 
the $1F_3'$ with $M=3218$~MeV and $\Gamma=323$~MeV, 
the $2D_2$ with $M=3298$~MeV and $\Gamma=106$~MeV 
and the $2D_2'$ with $M=3323$~MeV and $\Gamma=203$~MeV.  However we consider all of these
possibilities unlikely as the predicted masses are over 100~MeV from the observed mass.  And
although it would not surprise us if our predictions were off by several tens of MeV we 
do not expect them to be off by over 100~MeV.  A final possibility is the
$3^1S_0$ with $M=3154$~MeV and $\Gamma=79$~MeV with BR to $D^*K$ of 15\%.  In this case 
the predicted width is smaller by a factor of three so that it seems unlikely that the 
 $D_{sJ}(3044)^\pm$ could be identified as the $3^1S_0$.  
 
To summarize,  with the information we currently have for the $D_{sJ}(3044)^\pm$ it
is most likely either the $D_{s1}(2P_1^\prime)$ or the $D_{s1}(2P_1)$ or both states overlapping.
This conclusion is consistent with other studies 
\cite{Chen:2009zt,Sun:2009tg,Zhong:2009sk,Colangelo:2010te,Li:2010vx,Colangelo:2012xi,Xiao:2014ura}. 
Another possibility is that it is $D_{s2}^* (2^3P_2)$ with the signal for the $DK$ final
state too small to be observed with current statistics.  These different possibilities can
be tested by measuring BR's to $DK^*$ and $D_s\phi$ final states.  We also expect that 
it should be possible to observe the $2P$ partners which will lie in this mass region
in $DK$ and $D^*K^*$ final states.

\section{Finding the Missing Charm Mesons}
\label{sec:finding_the_missing}

The key to observing missing states is that their total width is not too large and that 
the BR's to at least some simple final states are not too small. This is how the new charm states
were found by the BaBar and LHCb collaborations.  Thus,  we can use our tables of charm
meson properties to identify candidate states that could be observed in the near future.
As the states become more massive, more and more channels open up so that the BR's to easier to 
observe final states become smaller and smaller. For masses above around 3500~MeV for charm
mesons  the BR's to simple
final states are less than 1\% and are likely too small
to observe.  For charm-strange mesons, BR's to at least some 
simple final states remain non negligible for all states we consider due to the
smaller phase space because of the larger kaon mass relative to that of the pion.  
Another consideration is that states within multiplets will be overlapping and states in
different multiplets are close enough in mass that it will require more than ``bump hunting'' 
to classify newly found states. Determining the spin of a state and measuring BR's to 
multiple final states will be important to disentangle the spectrum.  We have already
seen examples of this in the preceding sections.

\subsection{The Charm Mesons}

For the most part, the recently observed states are the states with large BR's to simple 
final states.  For 
example the predicted BR's of the $2^3S_1$ and $2^1S_0$  to the observed final state $D^*\pi$ 
are 58\% and 99\% respectively.
Likewise the $1D$ states have BR's to $D^*\pi$ ranging from 13\% to 38\% and the $1^3F_4$ has a BR to
$D\pi$ of 12\%.  We will use BR's to simple final states to identify good candidates for 
discovery.

We start with the $1D$ multiplet. Three of the states have been observed, the $1^3D_1$ and 
$1^3D_3$ and tentatively the $1D_2$ leaving only the $1D_2'$ to be found.  This state has a
BR of 38\% to $D^*\pi$ but is predicted to be rather broad, $\sim 240$~MeV, making it potentially
difficult to disentangle from the other three $1D$ states in that mass region.  This state 
might also be seen in the $D_s^*K$ final state.

We tentatively identified the $3^1S_0$ state with the $D_J(3000)$ although the $1F_3$
and $1F_3'$ are also possibilities.  If we accept the $3^1S_0$ assignment we would expect 
that the $3^3S_1$ should also be seen with comparable statistics.  The distinguishing 
feature is that the $3^3S_1$ should be seen in both $D\pi$ and $D^*\pi$ final states. Even
if the  $D_J(3000)$ turns out to be the $1F_3$ or $1F_3'$ we expect that the $3S$ states 
could be seen in the near future.

The $2P$ states also have relatively large BR's to $D^*\pi$ and $D\pi$ final states. 
Their masses are expected to be in the 2900-2950~MeV mass range with widths ranging from
114 to 212~MeV.  In fact, some have argued
that the $2^3P_0$ can be identified with the $D_J^*(3000)$.  We expect that
the $2P$ multiplet can be observed in $D^*\pi$ and $D\pi$ final states.  The four states
are only split by 37~MeV so that it will require the measurement into different final states
to uniquely identify the individual states.  As we pointed out previously, in addition
to $D^*\pi$ and $D\pi$, the  $D\rho$ final state will be a useful discriminator.  Other
final states which would help are the $D_sK$, $D_s^* K$, $D_s K^*$ and $D^*\rho$ although 
in some cases they only have a sizeable BR for one of the $2P$ states.  In this case their
observation in one of these final states would eliminate other possibilities.

The $1F$ multiplet is next in line using this criteria for ``discoverability'' with BR's to
$D\pi$ and $D^*\pi$ ranging from 8 to 20\%.  Their masses are around 3100~MeV with predicted
widths ranging from 126 to 270~MeV.  Depending on the reliability of our width predictions,
the two broad states, the $1^3F_2$ and $1F_3'$, are likely too broad to be easily 
seen.  We have tentatively identified the $1^3F_4$ with the $D_J^*(3000)$ state leaving the
$1F_3$ to be found.  If found, it is expected to have a large BR into $D\rho$ which 
could be used as confirmation.  

The $1G$ multiplet also has a significant BR to the $D\pi$ and $D^*\pi$ final states ranging 
from 4\% to 17\% depending on the state.  Their masses range from around 3360 to 3400~MeV and 
their widths range from 118 to 254~MeV.  The narrower widths correspond to the $j=9/2$ doublet
and the broader widths to the $j=7/2$ doublet.  We expect it more likely that the narrower 
$1^3G_5$ and $1G_4$ states will be observed first.  The natural parity $1^3G_5$ decays to both $D\pi$
and $D^*\pi$ while the un-natural $1G_4$ can only decay to $D^*\pi$.  Other decay modes that 
can be used to distinguish between these states are $D\rho$ where  
$BR(D(1^3G_5) \to D\rho)\simeq 3\%$ versus $BR(D(1G_4) \to D\rho)\simeq 15\%$
and $D^*\rho$ where $BR(D(1^3G_5) \to D^*\rho)\simeq 32\%$ versus $BR(D(1G_4) \to D^*\rho)\simeq 17\%$. 
One could use similar measurements to identify the $1G_4'$ and $1^3G_3$.

Beyond these multiplets,  the BR's to  $D\pi$ and $D^*\pi$ final states 
for the most part become relatively small and other final states will become more important for finding higher
excited missing states. We already suggested that the $D\rho$ and $D^*\rho$ final states would be
useful for identifying excited charm states and for many of the higher excited states they have the largest
BR's and could prove crucial for their discovery.   
For example, in the  $2D$ multiplet $BR(2^3D_1 \to D\pi) \simeq 5.1\%$ but 
$BR(2^3D_1 \to D^*\rho) \simeq 13.9\%$.  The challenge is that the $D^*$ and $\rho$ will have to be 
reconstructed with numerous pions in the final state.  Similarly,
$BR(2^3D_3 \to D^*\rho) \simeq 11.4\%$ versus $BR(2^3D_3 \to D^*\pi) \simeq 3.0\%$
and $BR(2D_2 \to D^*\rho) \simeq 14.7\%$ versus $BR(2D_2 \to D^*\pi) \simeq 9.5\%$.  To complete
the multiplet we note that 
$BR(2D_2' \to D^*\rho) \simeq 7.3\%$ versus $BR(2D_2^\prime \to D^*\pi) \simeq 7.8\%$ so that
the ratio $BR(2D_2^{(\prime)} \to D^*\rho)/BR(2D_2^{(\prime)} \to D^*\pi)$ could be useful for discriminating 
between $2D_2'$ and $2D_2$.  

One can continue this exercise by examining the predicted BR's of higher mass multiplets given in 
Tables~\ref{tab:D_4S1}-\ref{tab:D_2G3}. 
Further examples that satisfy the criteria of being not too broad while having
a not too small BR to a simple final state like $D\pi$ and $D^*\pi$ are members of
the $2D$ and $3P$ multiplets etc.  The interested reader can identify more candidate states that might be found
in the near future by examining Tables~\ref{tab:D_4S1}-\ref{tab:D_2G3}.

\subsection{The Charm-strange Mesons}

We will follow the approach used in the previous section 
to identify likely charm-strange discovery candidates 
 using the criteria that states with large branching ratios to simple final states are the ones
most likely to be observed.  

The charm-strange states that have recently been observed by Belle, BaBar and LHCb all follow this pattern
of large BR's to simple states.  For example, the $D^*_{s1}(2709)$ identified as the $D_s^*(2^3S_1)$
is predicted to have $BR(D_s^*(2^3S_1) \to D^*K)  \simeq 58.6\%$ and  
$BR(D_s^*(2^3S_1) \to DK)  \simeq 32.5\%$, the discovery channels.  Similarly the 
$D^*_{s1}(2860)$ and $D^*_{s3}(2860)$ are identified as the $D_s^*(1^3D_1)$ 
and $D_s^*(1^3D_3)$ which are predicted to have BR's to $DK$,
the channel studied by LHCb, of  47.3\% and 57.7\% respectively.  
Finally,  we tentatively identify the $D_{sJ}(3044)$ as either $2P_1$ or $2P_1'$ whose 
calculated BR's to $D^*K$, the BaBar discovery channel, are 42.9\% and 24.7\% respectively.
The common thread is that all of the recently discovered $D_{sJ}^{(*)}$ have large BR's to
simple final states.

If we extrapolate this to other excited states we can expect many more $D_{sJ}^{(*)}$
to be discovered in the near future.  The singlet partner of the $D^*_{s1}(2709)$, 
the $D_s(2^1S_0)$, 
is expected to lie 59~MeV lower at 2650~MeV with a width of 74~MeV 
and a BR of almost 100\% to $D^*K$.  The $J=2$ partners of $D^*_{s1}(2860)$ and $D^*_{s3}(2860)$ 
are the $1D_2$ and $1D_2'$, which should lie very close in mass to the $D^*_{s1}$ and
$D^*_{s3}$.  Their widths are expected to be 115 and 195~MeV respectively with BR's
to $D^*K$ of 18.3\% and 83.4\%.  Finally,  if the $D_{sJ}(3044)$  is the $2P_1$ or $2P_1'$
there will be three other $2P$ partner states,  the $2^3P_2$,  $2^3P_0$, and the other 
$J=1$ state.  Because we did not determine if the  $D_{sJ}(3044)$ is the $2P_1$ or $2P_1'$
we will make a few comments about all four of the $2P$ states and 
refer to Table~\ref{tab:Ds_2P1} for details.  All 
four of the $2P$ states are close in mass ranging from 3005~MeV for the $2^3P_0$ to 3048~MeV for
the $2^3P_2$ state.  The widths of all four states are around 130-150~MeV.  What distinguishes
them are their BR's to different final states: $D_s^*(2^3P_0)$ decays to $DK$ with 
$BR\simeq 35\%$ but does not decay to $D^*K$, while the $D_s(2P_1)$ and $D_s(2P_1')$ decay 
to $D^*K$ with $BR\simeq 42.9\%$ and 24.7\% respectively but do not decay to $DK$,
and  $D_s^*(2^3P_2)$ can decay to both $DK$ and $D^*K$ with BR's of 6.5\% and 23.4\% 
respectively. The point is that all of these states have sizeable BR's to final states
that have already lead to the observation of a new state in this mass region so that we 
expect that the remaining three states should also be seen.

So far we have only discussed charm-strange mesons that are members of multiplets with states 
that have already been
observed.  The next most promising states to find are likely to be members of the $1F$
multiplet and in fact the $1F_3$ states are alternative possibilities for the $D_{sJ}(3044)$. 
We leaned towards the $2P_1$ identification primarily because the predicted mass was closer 
to the observed mass.  In any case the $1F$ states are expected to sit around
3200~MeV.  The $1^3F_4$ is predicted to have a width of 182~MeV with BR's to $DK$ and $D^*K$
of 15.9\% and 13.6\% respectively, the $1F_3$ and $1F_3'$ have widths of 183~MeV and 323~MeV respectively
with BR's to $D^*K$ of 22.6\% and 28.9\% respectively and do not decay to $DK$, and the $1^3F_2$ 
is predicted to have a width of  292~MeV and decays to $DK$ and $D^*K$
with BR's of 15.3\% and 12.9\% respectively.  Although the BR's are all sizeable, 
the $1F$ states are expected to be relatively broad so it is not clear if they will be 
observed in the near future.  However, as we have repeatedly pointed out, our width predictions
can easily be off by up to a factor of two so that if they turn out to be narrower than we
predict, their observation would be more likely.  

We note that the $1G$ multiplet has similar BR's to final states (see Table~\ref{tab:Ds_1G1})
as the $1F$ multiplet 
so it might also be possible to observe these states with the same caveat regarding their 
large total widths.  It would be extremely interesting to find these states as we would 
then have a series of angular momentum states stretching from $L=0$ to $L=4$ 
which would test 
the linearity of the Regge trajectory and thus the linearity of the confining potential
\cite{Godfrey:1985by}.  If large deviations were found it would also provide some insights
into the importance of meson loop contributions to the mass of excited states.  

Beyond these states there are a smattering of states that
have large BR's to the $DK$ and $D^*K$ final states we have focused on. A few 
examples are: 
the $3^3P_0$ with a predicted mass of 3412~MeV, width of 104~MeV and BR to $DK$ of 19.5\%, 
the $4^3P_0$ with $M=3764$~MeV, $\Gamma = 105$~MeV and BR to $DK$ of 10\%,
the $2D_2$ with $M=3298$~MeV, $\Gamma = 106$~MeV and BR to $D^*K$ of 11.4\%
and the $2D_2'$ with $M=3323$~MeV, $\Gamma = 203$~MeV and BR to $D^*K$ of 21.4\%. 
One can turn to Tables~\ref{tab:Ds_3P2}-\ref{tab:Ds_2G3}  to explore further possibilities.
As more measurements are made we will be able to gauge the reliability of our predictions and 
those of others and refine the models to improve their predictive power.

\section{Summary}

In this paper we calculated the properties of charm and charm-strange mesons using the
relativized quark model to calculate masses and wavefunctions which were used to calculate
radiative transition partial widths.  We calculated hadronic widths using the quark pair creation 
model with simple harmonic oscillator wavefunctions with the oscillator parameters fitted 
to the rms radius of the relativized quark model wavefunctions.  

We used our results to identify recently observed charm and charm-strange mesons in terms
of  quark model spectroscopic states.  Our results 
support the previously made assignment of the $D_J(2550)^0$ and $D_J^*(2600)^0$ as the 
$2^1S_0(c\bar{q})$  and $2^3S_1(c\bar{q})$ states respectively.  We 
identify the $D_1^*(2760)^0$ and $D_3^*(2760)^0$ as the
$1^3D_1(c\bar{q})$ and $1^3D_3(c\bar{q})$ respectively and tentatively identify the 
$D_J(2750)^0$ as the  $1D_2(c\bar{q})$ state.  In the latter case further measurements
are needed to strengthen the assignment.  We suggested that measurements of BR's to $D\rho$ 
and $D^*\pi$ would be useful.  We tentatively identified the $D_J^*(3000)^0$  as 
the $D_4^*(1^3F_4)$ state and favour the $D_J(3000)^0$ to be the $D(3^1S_0)$ although
we do not rule out the $1F_3$ and $1F_3'$ assignments.  
For the recently observed charm-strange mesons we identify the
$D_{s1}^*(2709)^\pm$, $D_{s1}^*(2860)^-$, and $D_{s3}^*(2860)^-$ as
the $2^3S_1(c\bar{s})$, $1^3D_1(s\bar{c})$, and $1^3D_3(s\bar{c})$ respectively and suggest
that the $D_{sJ}(3044)^\pm$ is most likely the $D_{s1}(2P_1^\prime)$ or  $D_{s1}(2P_1)$
although it might be the $D_{s2}^*(2^3P_2)$ with the $DK$ final state  too small
to be observed with current statistics.  

Finally we suggested excited charm and charm-strange mesons that might be seen in the near future
based on the criteria that they do not have too large a total width and they have a reasonable branching 
ratio to simple final states.  We expect that our results comprised of tables of masses, widths
and BR's will be useful to this end.    

While we have shown the usefulness of our results in identifying newly discovered states we are equally
keen that they  be a useful guide for future searches for missing states.

\acknowledgments

This research was supported in part by 
the Natural Sciences and Engineering Research Council of Canada under grant number 121209-2009 SAPIN.


\end{document}